# THERMODYNAMICS OF MIXTURES CONTAINING A VERY STRONGLY POLAR COMPOUND. 11. 1-ALKANOL + ALKANENITRILE SYSTEMS

JUAN ANTONIO GONZÁLEZ*(1), FERNANDO HEVIA(1), ANA COBOS(1), ISAÍAS GARCÍA DE LA FUENTE(1), AND CRISTINA ALONSO TRISTÁN(2)

(1) G.E.T.E.F., Departamento de Física Aplicada, Facultad de Ciencias, Universidad de Valladolid, Paseo de Belén, 7, 47011 Valladolid, Spain,

(2) Dpto. Ingeniería Electromecánica. Escuela Politécnica Superior. Avda. Cantabria s/n. 09006 Burgos, (Spain)

*e-mail: jagl@termo.uva.es; Fax: +34-983-423136; Tel: +34-983-423757

**ABSTRACT**

1-Alkanol + alkanenitrile systems have been studied by means of the DISQUAC, ERAS and UNIFAC (Dortmund) models. DISQUAC and ERAS parameters for the alkanol/nitrile interactions are reported. DISQUAC describes a whole set of thermodynamic properties: phase equilibria, vapour-liquid (VLE), liquid-liquid (LLE) and solid-liquid (SLE) equilibria, molar excess functions, Gibbs energies, $G_{\mathrm{m}}^{\mathrm{E}}$, and enthalpies, $H_{\mathrm{m}}^{\mathrm{E}}$, and partial excess molar enthalpies at infinite dilution, $H_{\mathrm{mi}}^{\mathrm{E},\infty}$ using the same set of interaction parameters for each solution. The dependence on the molecular structure of the interaction parameters is similar to that observed in other previous applications to mixtures formed by 1-alkanols and a strongly polar compound, in such way that the quasichemical interchange coefficients can be kept constant from 1-propanol. However, methanol and ethanol solutions behave differently. From the analysis of experimental data for $H_{\mathrm{m}}^{\mathrm{E}}$, $TS_{\mathrm{m}}^{\mathrm{E}}(=H_{\mathrm{m}}^{\mathrm{E}}-G_{\mathrm{m}}^{\mathrm{E}})$, and molar excess volumes, $V_{\mathrm{m}}^{\mathrm{E}}$, it is concluded that the studied systems are characterized by dipolar interactions and strong structural effects. The former are more relevant in acetonitrile solutions. Association effects are more important in butanenitrile mixtures. DISQUAC improves $H_{\mathrm{m}}^{\mathrm{E}}$ results from the ERAS model. ERAS results on $H_{\mathrm{m}}^{\mathrm{E}}$ for systems containing acetonitrile are also improved by UNIFAC. This remarks the importance of dipolar interactions in the investigated mixtures. ERAS describes the variation of $V_{\mathrm{m}}^{\mathrm{E}}$ ( $x_1$ =0.5) with the 1-alkanol size for mixtures with a given nitrile, but the concentration dependence of this excess function is poorly represented.

## 1. INTRODUCTION

Nitriles are important chemicals in the production of plastics, cosmetics or pharmaceuticals [1]. Nitriles are also good hydrogen bond acceptors [2,3]. For example, many hydrogen bonds are between the nitrile and aminoacids such as serine or arginine [1]. On the other hand, the high polarity of nitriles allows them to act as hydroxyl or carboxyl isosteres. The acetonitrile + water system is used as mobile phases in reversed phase high performance liquid chromatography, which is commonly the final step in protein purification [4].

1-Alkanol + alkanenitrile mixtures have been studied in terms of a variety of theories. Systems formed by methanol, ethanol, 1-propanol or 1-butanol and acetonitrile have been investigated using different association models including physical interactions by means of the UNIQUAC or NRTL equations [5-10]. The latter model assumes the existence of linear and cyclic species for alcohols and acetonitrile, and only open chains for alcohol-nitrile complexes [5-7]. The association theory what includes the UNIQUAC equation is applied taking into consideration only linear polymers for 1-alcohol self-association and for alcohol-nitrile solvation [8-10]. The ERAS model [11] has been used for the correlation of molar excess volumes of this type of systems assuming that acetonitrile is either a self-associated [12] or a non self-associated [13] compound. However, there are very limited studies on the simultaneous representation of both $H_{\mathrm{m}}^{\mathrm{E}}$ and $V_{\mathrm{m}}^{\mathrm{E}}$ of 1-alkanol + nitrile mixtures [14,15]. Interaction parameters for the hydroxyl/nitrile interactions are available [16] in the framework of UNIFAC (Dortmund version) [17]. In order to gain a deeper insight into the interactions and structure of these solutions, we apply now the DISQUAC [18] and ERAS models for the characterization of 1-alkanol + alkanenitrile mixtures. Interestingly, alkanenitriles are compounds with very high dipole moments (e.g., 3.53 D and 3.50 D for acetonitrile or butanenitrile, respectively [19]), and 1-octanol or 1-decanol + acetonitrile systems show miscibility gaps at temperatures close to 298.15 K. In fact, the upper critical solution temperature (UCST) of such mixtures is 282.7 and 297.05 K, respectively [15]. Thus, one can expect that the studied systems show strong positive deviations from the Raoult's law. Previously, we have investigated other solutions formed by 1-alkanols and a highly polar compound, as sulfolane [20], amides [21-23] or dimethylsulfoxide [24].

## 2. MODELS

### *2.1 DISQUAC*

The DISQUAC group contribution model [18] is based on the rigid lattice theory developed by Guggenheim [25]. Some important features of the model are the following. (i) The total molecular volumes, $r_{\mathrm{i}}$, surfaces, $q_{\mathrm{i}}$, and the molecular surface fractions, $\alpha_{\mathrm{si}}$, of the compounds present in the solution are calculated additively using the group volumes $R_{\mathrm{G}}$ and surfaces $Q_{\mathrm{G}}$ recommended by Bondi [26]. As volume and surface units, the volume $R_{\mathrm{CH4}}$ and surface $Q_{CH4}$ of methane are taken arbitrarily [27]. The geometrical parameters for the groups referred to in this work are available in the literature [27-29] (ii) The partition function is factorized into two terms, and the excess functions are calculated as the sum of two contributions. The dispersive (DIS) term represents the contribution from the dispersive interactions. The quasichemical (QUAC) term is related to the anisotropy of the field forces created by the solution molecules. In the case of the Gibbs energy, $G_{\mathrm{m}}^{\mathrm{E}}$, a combinatorial term, $G_{\mathrm{m}}^{\mathrm{E,COMB}}$, represented by the Flory-Huggins equation [27,30] must be also taken into account. Thus,

$$G_{\mathrm{m}}^{\mathrm{E}} = G_{\mathrm{m}}^{\mathrm{E,DIS}} + G_{\mathrm{m}}^{\mathrm{E,QUAC}} + G_{\mathrm{m}}^{\mathrm{E,COMB}} \qquad (1)$$

$$H_{\mathrm{m}}^{\mathrm{E}} = H_{\mathrm{m}}^{\mathrm{E,DIS}} + H_{\mathrm{m}}^{\mathrm{E,QUAC}} \qquad (2)$$

(iii) It is assumed that the interaction parameters depend on the molecular structure; (iv) The value $z$ = 4 for the coordination number is used for all the polar contacts. This is an important shortcoming of the model, partially removed via the hypothesis of considering structure dependent interaction parameters. (v) The mixing process takes place without change in volume.

The equations used to calculate the DIS and QUAC contributions to $G_{\mathrm{m}}^{\mathrm{E}}$ and $H_{\mathrm{m}}^{\mathrm{E}}$ in the framework of DISQUAC are given elsewhere [28,31]. The temperature dependence of the interaction parameters is expressed in terms of the DIS and QUAC interchange coefficients [31], $C_{\mathrm{st},l}^{\mathrm{DIS}}; C_{\mathrm{st},l}^{\mathrm{QUAC}}$ where s ≠ t are two contact surfaces present in the mixture and $l$ = 1 (Gibbs energy; $C_{\mathrm{st},1}^{\mathrm{DIS/QUAC}} = g_{\mathrm{st}}^{\mathrm{DIS/QUAC}}(T_{\mathrm{o}})/RT_{\mathrm{o}}$); $l$ = 2 (enthalpy, $C_{\mathrm{st},2}^{\mathrm{DIS/QUAC}} = h_{\mathrm{st}}^{\mathrm{DIS/QUAC}}(T_{\mathrm{o}})/RT_{\mathrm{o}}$)), $l$ = 3 (heat capacity, $C_{\mathrm{st},3}^{\mathrm{DIS/QUAC}} = c_{\mathrm{pst}}^{\mathrm{DIS/QUAC}}(T_{\mathrm{o}})/R$)). $T_{\mathrm{o}}$ = 298.15 K is the scaling temperature and $R$, the gas constant. The equations are [31]:

$$g_{\mathrm{st}}^{\mathrm{DIS/QUAC}}/RT = C_{\mathrm{st},1}^{\mathrm{DIS/QUAC}} + C_{\mathrm{st},2}^{\mathrm{DIS/QUAC}}\left(\frac{T_o}{T}-1\right) + C_{\mathrm{st},3}^{\mathrm{DIS/QUAC}}\left[\ln\left(\frac{T_o}{T}\right) - \left(\frac{T_o}{T}\right) + 1\right] \qquad (3)$$

$$h_{\mathrm{st}}^{\mathrm{DIS/QUAC}}/RT = C_{\mathrm{st},2}^{\mathrm{DIS/QUAC}}\left(\frac{T_o}{T}\right) - C_{\mathrm{st},3}^{\mathrm{DIS/QUAC}}\left[\left(\frac{T_o}{T}\right) - 1\right] \qquad (4)$$

$$c_{\mathrm{p,st}}^{\mathrm{DIS/QUAC}}/R = C_{\mathrm{st},3}^{\mathrm{DIS/QUAC}} \qquad (5)$$

DISQUAC calculations for the coexistence curves for the LLE curves were conducted taking into account that the values of the mole fraction $x_1$ of component 1 ($x_1^{'}, x_1^{''}$) relating to the two phases in equilibrium are such that the functions $G_{\mathrm{m}}^{\mathrm{M'}}, G_{m}^{M''}$ ($G_{\mathrm{m}}^{\mathrm{M}} = G_{\mathrm{m}}^{\mathrm{E}} + G_{\mathrm{m}}^{\mathrm{ideal}}$) have a common tangent [32]. On the other hand, the SLE curve of a pure solid component 1 is represented by the expression [33]:

$$-\ln x_1 = \frac{\Delta H_{m1}}{R}\left[\frac{1}{T}-\frac{1}{T_{m1}}\right]-\frac{\Delta C_{Pm1}}{R}\left[\ln(\frac{T}{T_{m1}})+\frac{T_{m1}}{T}-1\right]+\ln\gamma_1 \quad (6)$$

Conditions at which eq. (6) is valid have been specified in previous works [30]. In eq. (6), $x_1$ is the mole fraction and $\gamma_1$ the activity coefficient of component 1 in the solvent mixture, at temperature *T*. In this work, DISQUAC is used to calculate $\gamma_1$. The remaining symbols $\Delta H_{m1}, T_{m1}, \Delta C_{Pm1}$ have their usual meaning [15]. Values needed for calculations were taken from [15].

*2.2 Modified UNIFAC (Dortmund version)*

Modified UNIFAC [16,17] differs from the original UNIFAC [34] by the combinatorial term and the temperature dependence of the interaction parameters. The equations used to calculate $G_{\mathrm{m}}^{\mathrm{E}}$ and $H_{\mathrm{m}}^{\mathrm{E}}$ are obtained from the fundamental equation for the activity coefficient $\gamma_i$ of component i:

$$\ln\gamma_i = \ln\gamma_i^{COMB} + \ln\gamma_i^{RES} \quad (7)$$

where $\ln\gamma_i^{COMB}$ and $\ln\gamma_i^{RES}$ stand for the combinatorial and residual term, respectively. Equations can be found elsewhere [31]. In this version of UNIFAC, two main groups, OH and $CH_3OH$, are defined for the prediction of thermodynamic properties of mixtures containing alkanols. The main group OH is subdivided in three subgroups: OH(p), OH(s) and OH(t) for the representation of primary, secondary and tertiary alkanols, respectively. The $CH_3OH$ group is a specific group for methanol solutions. Alkanenitriles are represented by only one main group, CCN, subdivided in two subgroups, $CH_3CN$ and $CH_2CNH_2$. The subgroups have different geometrical parameters, but the subgroups within the same main group are assumed to have identical group energy-interaction parameters. On the other hand, the geometrical parameters, the relative van der Waals volumes and the relative van der Waals surfaces are not calculated form molecular parameters like in the original UNIFAC, but fitted together with the interaction parameters to the experimental values of the thermodynamic properties considered. The geometrical and interaction parameters were taken from literature and used without modifications [16].

*2.3 ERAS*

In the ERAS model [11], the excess functions are determined as the sum of two contributions. The chemical contribution, $X^{\mathrm{E}}_{\mathrm{m,chem}}$, arises from hydrogen-bonding effects. The physical contribution, $X^{\mathrm{E}}_{\mathrm{m,phys}}$, is related to non-polar van der Waals' interactions and free volume effects. The excess functions are written as:

$$X^{\mathrm{E}}_{\mathrm{m}} = X^{\mathrm{E}}_{\mathrm{m,phys}} + X^{\mathrm{E}}_{\mathrm{m,chem}} \qquad (8)$$

where $X^{\mathrm{E}}_{\mathrm{m}} = H^{\mathrm{E}}_{\mathrm{m}},\ V^{\mathrm{E}}_{\mathrm{m}},\ G^{\mathrm{E}}_{\mathrm{m}}$. The resulting expressions for these functions are given elsewhere [35]. In addition, only consecutive linear association is assumed, which is described by a chemical equilibrium constant ($K_{\mathrm{A}}$) independent of the chain length of the associated species (1-alkanols), according to the equation:

$$\mathrm{A_m} + \mathrm{A} \leftrightarrow \mathrm{A_{m+1}} \qquad (9)$$

with *m* ranging from 1 to $\infty$. The cross-association between a self-associated species $A_{\mathrm{m}}$ (1-alkanols) and a non self-associated compound *B* (nitriles) is represented by

$$A_{\mathrm{m}} + B \xleftrightarrow{K_{\mathrm{AB}}} A_{\mathrm{m}}B \qquad (10)$$

The association constants ($K_{\mathrm{AB}}$) of equation (10) are also assumed to be independent of the chain length. Equations (9) and (10) are characterized by $\Delta h^*_{\mathrm{i}}$, the enthalpy of the reaction that corresponds to the hydrogen-bonding energy, and by the volume change ($\Delta v^*_{\mathrm{i}}$) related to the formation of the linear chains.

$X^{\mathrm{E}}_{\mathrm{m,phys}}$ is derived from the Flory's equation of state [36], which is assumed to be valid not only for pure compounds but also for the mixture [37,38].

$$\frac{\bar{P}_{\mathrm{i}}\bar{V}_{\mathrm{i}}}{\bar{T}_{\mathrm{i}}} = \frac{\bar{V}_{\mathrm{i}}^{1/3}}{\bar{V}_{\mathrm{i}}^{1/3}-1} - \frac{1}{\bar{V}_{\mathrm{i}}\,\bar{T}_{\mathrm{i}}} \qquad (11)$$

where i = A,B or M (mixture). In equation (11), $\bar{V}_{\mathrm{i}} = V_{\mathrm{i}}/V^*_{\mathrm{i}}$; $\bar{P}_{\mathrm{i}} = P/P^*_{\mathrm{i}}$; $\bar{T}_{\mathrm{i}} = T/T^*_{\mathrm{i}}$ are the reduced volume, pressure and temperature respectively. The pure component reduction parameters $V_i^*$, $P_i^*$, $T_i^*$ are determined from *P*-*V*-*T* data (density, $\rho$, isobaric expansion coefficient, $\alpha_p$, and isothermal compressibility, $\kappa_T$), and association parameters [37,38]. The reduction parameters for the mixture $P_M^*$ and $T_M^*$ are calculated from mixing rules [37,38]. The total relative molecular volumes and surfaces of the compounds were calculated additively using the Bondi's method (see above) [26].

## 3. ADJUSTMENT OF MODEL PARAMETERS

### *3.1 DISQUAC*

In the framework of DISQUAC, 1-alkanol + alkanenitrile mixtures are regarded as possessing the following three types of surface: (i) type a, aliphatic ($CH_3$, $CH_2$, in 1-alkanols and nitriles); (ii) type h, hydroxyl (OH in 1-alkanols); (iii) type n, (CN in alkanenitriles)

The investigated solutions are built by three contacts: (a,h), (a,n), and (h,n). The DIS and QUAC interaction parameters for the (a,h) and (a,n) contacts are known from the study of 1-alkanol [28,39] or alkanenitrile [29] + *n*-alkane mixtures, respectively. Thus, only the DIS and QUAC interaction parameters of the (k,n) contacts must be fitted to VLE and $H_m^E$ data of the studied systems. The general procedure applied in the estimation of the interaction parameters has been explained in detail elsewhere [22,31]. Final values of the fitted parameters in this work are collected in Table 1.

### *3.2 ERAS*

Values of $V_i$, $V_i^*$ and $P_i^*$ at $T$ = 298.15 K of 1-alkanols have been taken from a previous work [40]. At 298.15 K, such parameters for acetonitrile are: $V_i$ = 52.87 $cm^3 \cdot mol^{-1}$; $V_i^*$ = 40.20 $cm^3 \cdot mol^{-1}$; $P_i^*$ = 650.8 $J \cdot cm^{-3}$ (values determined using $\rho, \alpha_p$, and $\kappa_T$ data from [19]); and for butanenitrile, $V_i$ = 87.86 $cm^3 \cdot mol^{-1}$; $V_i^*$ = 68.56 $cm^3 \cdot mol^{-1}$; $P_i^*$ = 578.1 $J \cdot cm^{-3}$ values determined using $\rho, \alpha_p$, and $\kappa_T$ data from [19,41]). In the present case, as cross-association between components exists, the binary parameters to be fitted to the $H_m^E$ and $V_m^E$ data available in the literature for 1-alkanol + alkanenitrile mixtures are $K_{AB}$, $\Delta h_{AB}^*$, $\Delta v_{AB}^*$ and $X_{AB}$. $K_A$, $\Delta h_A^*$ (= −25.1 $kJ \cdot mol^{-1}$) and $\Delta v_A^*$ (= −5.6 $cm^3 \cdot mol^{-1}$) for 1-alkanols are known from $H_m^E$ and $V_m^E$ data for the corresponding systems with alkanes [37,38,40,42]. Adjusted ERAS parameters for the binary mixtures investigated are collected in Table 2.

## 4. THEORETICAL RESULTS

Theoretical results from DISQUAC and UNIFAC models for VLE, $G_{\mathrm{m}}^{\mathrm{E}}$ and $H_{\mathrm{m}}^{\mathrm{E}}$ are listed in Tables 3 and 4. For the sake of clarity, relative deviations for the pressure (*P*) and $H_{\mathrm{m}}^{\mathrm{E}}$ defined as:

$$\sigma_r(P) = \{\frac{1}{N}\sum\left[\frac{P_{\mathrm{exp}} - P_{\mathrm{calc}}}{P_{\mathrm{exp}}}\right]^2\}^{1/2} \quad (12)$$

and

$$dev(H_{\mathrm{m}}^{\mathrm{E}}) = \{\frac{1}{N}\sum\left[\frac{H_{\mathrm{m,exp}}^{\mathrm{E}} - H_{\mathrm{m,calc}}^{\mathrm{E}}}{H_{\mathrm{m,exp}}^{\mathrm{E}}(x_1 = 0.5)}\right]^2\}^{1/2} \quad (13)$$

are included in Tables 3 and 4, respectively. DISQUAC results for SLE are collected in Table 5. Deviations for the equilibrium temperatures, listed in Table 5, are calculated from the equations:

$$\sigma_r(T) = \{\frac{1}{N}\sum\left[\frac{T_{\mathrm{exp}} - T_{\mathrm{calc}}}{T_{\mathrm{exp}}}\right]^2\}^{1/2} \quad (14)$$

$$\Delta(T)/K = \frac{1}{N}\sum\left|T_{\mathrm{exp}} - T_{calc}\right| \quad (15)$$

Table 6 compares experimental coordinates of the critical points for 1-octanol or 1-decanol + acetonitrile mixtures with values determined from DISQUAC. Table 7 shows a comparison between experimental measurements and DISQUAC calculations for partial excess molar enthalpies at infinite dilution, $H_{\mathrm{mi}}^{\mathrm{E},\infty}$. ERAS results on $H_{\mathrm{m}}^{\mathrm{E}}$ and $V_{\mathrm{m}}^{\mathrm{E}}$ are shown in Tables 4 and 8. Theoretical calculations for different thermodynamic properties are shown graphically, for some selected systems, along Figures 1-6. In view of the present results, one can conclude that DISQUAC describes properly VLE, SLE, LLE, $G_{\mathrm{m}}^{\mathrm{E}}$ and $H_{\mathrm{m}}^{\mathrm{E}}$ of 1-alkanol + alkanenitrile mixtures over a wide range of temperature.

## 5. DISCUSSION

$H_{\mathrm{m}}^{\mathrm{E}}$ values of 1-alkanol + alkanenitrile mixtures are large and positive (Table 4), indicating that the main contributions to this excess function come from the disruption of interactions between like molecules. Interestingly, $TS_{\mathrm{m}}^{\mathrm{E}} (= H_{\mathrm{m}}^{\mathrm{E}} - G_{\mathrm{m}}^{\mathrm{E}})$ values are also large and positive. For butanetrile solutions, $TS_{\mathrm{m}}^{\mathrm{E}}$/J·mol$^{-1}$ = 245 (methanol); 542 (ethanol); 768 (1-propanol); 908 (1-hexanol); 1006 (1-octanol); 1122 (1-decanol) (values determined from those listed in Tables 3 and 4 for $G_\mathrm{m}^\mathrm{E}$ and $H_\mathrm{m}^\mathrm{E}$). In addition, $\frac{\Delta H_m^E}{\Delta T}$ values of methanol or ethanol + acetonitrile mixtures are 5.3 and 7.4 J·mol$^{-1}$·K$^{-1}$, respectively [5]. On the basis of such features, one can conclude that interactions in the investigated systems are essentially of dipolar type. 1-Alkanol + alkane mixtures, characterized by alcohol self-association, show large and positive isobaric excess heat capacities, $C_\mathrm{p,m}^\mathrm{E}$, (11.7 J·mol$^{-1}$·K$^{-1}$ for ethanol + heptane [43]), and large and negative $TS_\mathrm{m}^\mathrm{E}$ values. For the ethanol + hexane system, $H_\mathrm{m}^\mathrm{E}$ = 548 [44]; $G_\mathrm{m}^\mathrm{E}$ = 1374 [45] and $TS_\mathrm{m}^\mathrm{E} = -826$ (data in J·mol$^{-1}$). On the other hand, $H_\mathrm{m}^\mathrm{E}$ values of mixtures containing, say, heptane, are rather low and increase from ethanol to 1-propanol or 1-butanol and then smoothly decrease. It is remarkable that the $H_\mathrm{m}^\mathrm{E}$ curves are shifted to low mole fractions of the alcohol [28,40,44]. We note that $H_\mathrm{m}^\mathrm{E}$ values of mixtures with a given nitrile increase with the chain length of the alcohol (Table 4). This can be ascribed to: (i) alcohols with larger aliphatic surfaces break more easily interactions between nitrile molecules. Accordingly, $H_\mathrm{m}^\mathrm{E}$ values of nitrile + alkane mixtures increase with the alkane size. For example, $H_\mathrm{m}^\mathrm{E}$(butanenitrile)/J·mol$^{-1}$= 1304 (heptane) [46], 1554 (dodecane) [47]; 1702 (tetradecane) [47]. (ii) Interactions between unlike molecules become less probable for longer alcohols as the hydroxyl group is more sterically hindered. The replacement of acetonitrile by butanenitrile in mixtures with a given 1-alkanol leads to decreased $H_\mathrm{m}^\mathrm{E}$ values (Table 4). This may be due to the positive contribution to $H_\mathrm{m}^\mathrm{E}$ from the breaking of nitrile-nitrile interactions is lower in the case of butanenitrile mixtures. In fact, dipolar interactions between acetonitrile molecules are stronger, as it is indicated by: (i) the existence of miscibility gaps at temperatures close to 298.15 K for systems formed by this nitrile and long chain 1-alkanols (Table 6, [15]); (ii) the much higher upper critical solutions temperatures (UCST) of acetonitrile + alkane mixtures compared to those of butanenitrile solutions. Thus, UCST(hexane)/K = 350.2 (acetonitrile) > 244.2 (butanenitrile) [48]. Many other systems show a similar behavior. Some examples follow: $H_\mathrm{m}^\mathrm{E}$(*N,N*-dimethylformide)/J·mol$^{-1}$= $-103$ (methanol) [49]; 968 (1-butanol) [49]; $H_\mathrm{m}^\mathrm{E}$(*N,N*-dimethylacetamide)/J·mol$^{-1}$= $-737$ (methanol) [50]; 367 (1-butanol) [51]; $H_\mathrm{m}^\mathrm{E}$(1,4-dioxane)/J·mol$^{-1}$ = 1134 (methanol) [52]; 1979 (1-butanol) [53]; $H_\mathrm{m}^\mathrm{E}$(tetrahydropyran)/J·mol$^{-1}$= 683 (methanol) [52]; 922 (1-butanol) [54]; or $H_\mathrm{m}^\mathrm{E}$(1-butanol)/J·mol$^{-1}$ = 2356 (dimethyl carbonate) [55]; 1944 (diethyl carbonate) [56].

From the application of the ERAS model, some interesting statements can be provided. i) The large $X_{AB}$ values and the low $K_{AB}$ values obtained for systems with 1-alkanols ≥ 1-propanol remark the importance of physical interactions in these solutions. Note that, as a trend, DISQUAC meaningfully improves ERAS results on $H_m^E$ (Table 4). (ii) The replacement of acetonitrile by butanenitrile in systems with a given 1-alkanol ($\neq$ methanol) leads to much lower $X_{AB}$ values (Table 2). This may be interpreted as a proof that physical interactions are more relevant in acetonitrile mixtures and that association effects become more relevant in systems with butanenitrile. Interestingly, $H_m^E$ results for acetonitrile solutions from UNIFAC are also better than those provided by ERAS (Table 4). (iii) The large $|\Delta v_{AB}^*|$ values obtained for systems with 1-alkanols $\geq$ 1-propanol (Table 2) point out to the existence of strong structural effects. This is supported by the lower $V_m^E$ values which contrast with the corresponding large positive $H_m^E$ data. Thus for the 1-butanol + acetonitrile system, $V_m^E$ /cm$^3$·mol$^{-1}$ = 0.1045 [57]; $H_m^E$/ J·mol$^{-1}$ = 2039 [7].

ERAS describes correctly the relative variation of $V_m^E$ (Table 8), but fails when representing the $V_m^E$ curves. Our results are similar to those provided in previous works when applying ERAS only to the description of $V_m^E$ .

Comparison of our ERAS parameters for the 1-alkanol-nitrile interactions with those available in the literature becomes rather difficult as different values for the $\Delta h_A^*$ and $\Delta v_A^*$ parameters of 1-alkanols were used [14,58]. For example, in a previous ERAS treatment of 1-nonanol, or 1-decanol + acetonitrile systems [58], the values of $\Delta h_A^*$ /kJ·mol$^{-1}$ = − 18.57 and of $\Delta v_A^*$ /cm$^3$·mol$^{-1}$= − 5.6 were used for 1-decanol, reporting for the mixture: $K_{AB}$ = 0.3; $\Delta h_{AB}^*$ /kJ·mol$^{-1}$ = − 17.1, $\Delta v_{AB}^*$ /cm$^3$·mol$^{-1}$ = − 1.2; $X_{AB}$ /J·cm$^{-3}$ = 146 and $dev(H_m^E)$ = 0.091. The model predicts $V_m^E$ = 1.89 cm$^3$·mol$^{-1}$, which is far from the experimental value (0.340 cm$^3$·mol$^{-1}$ [59]). That is, structural effects are not considered.

In terms of DISQUAC, we must remark that the interaction parameters for the (h,n) contacts in methanol or ethanol solutions are very different to those of the remaining mixtures. Along our investigations on 1-alkanol + alkanone [60], + linear alkanoate [61], + linear organic carbonate [62], + cyclic ether [63], + alkoxyethanol [64], + dimethylsulfoxide [24], + amide [21,22], or + amine [31,65] mixtures, we have shown that at least the $C_{sh,l}^{QUAC}$ (l=1,3) coefficients, and often also the $C_{sh,2}^{QUAC}$ parameter, can be kept constant along each homologous series. Thus, the present dependence on the molecular structure of the interaction parameters is rather anomalous. This might be related to the different behaviour of methanol or ethanol solutions, but also to certain experimental inaccuracies. Note the very different results available in the literature for ethanol or 1-propanol + butanenitrile systems (Table 4).

As usually, the model describes consistently a whole set of thermodynamic properties, VLE, SLE, LLE, $G_{\mathrm{m}}^{\mathrm{E}}$ and $H_{\mathrm{m}}^{\mathrm{E}}$ of a given solution using the same set of interaction parameters. The theoretical results obtained for the simultaneous representation of SLE and LLE of 1-octanol or 1-decanol + acetonitrile mixtures (Figure 6) must be underlined. Similar results have been obtained for, e.g, crown ether + alkane systems [66]. Deviations between experimental and theoretical results on LLE (Figure 6) are typically encountered when mean field theories, as DISQUAC, are used. In fact thermodynamic properties close to the critical points are expressed in terms of power laws, while for LLE, DISQUAC calculations are developed assuming erroneously that $G_{\mathrm{m}}^{\mathrm{E}}$ is an analytical function close to the critical point. The instability of a system is given by $(\partial^2 G_{\mathrm{m}}^{\mathrm{M}} / \partial x_1^2)_{P,T}$ and represented by the critical exponent $\gamma > 1$ in the critical exponents theory [67]. According to this theory, mean field models ($\gamma = 1$) provide LLE curves which are too high at the UCST and too low at the LCST [67] (lower critical solution temperature). For this reason, the $C_{\mathrm{ak,1}}^{\mathrm{DIS/QUAC}}$ coefficients must be ranged between certain limits in order to provide not very high calculated UCSTs. On the other hand, in mean field theories, $\beta = 0.5$ and LLE curves are more rounded close to the UCST (Figure 6), as fluctuations of the order parameter due to the rapid increase of the correlation length are not taken into account. The same trends are observed e.g., for sulfolane [24], crown ether [66], amide [21,22] or pyridine [65] + alkane mixtures. The larger deviations obtained for $H_{\mathrm{m}}^{\mathrm{E}}$ of 1-nonanol or 1-decanol + acetonitrile mixtures (Table 4) may be due, at least in part, to the proximity of the critical point.

## 6. CONCLUSIONS

The DISQUAC, UNIFAC and ERAS models have been applied to 1-alkanol + alkanenitrile mixtures. The DISQUAC and ERAS parameters for the alkanol/nitrile interactions are reported. The studied systems are characterized by dipolar interactions and strong structural effects. The former are more relevant in acetonitrile solutions. Association effects are more important in butanenitrile mixtures. DISQUAC improves $H_{\mathrm{m}}^{\mathrm{E}}$ results from the ERAS model. For mixtures containing acetonitrile, UNIFAC also improves such ERAS results. This remarks the importance of dipolar interactions in the investigated mixtures.

TABLE 1

Dispersive (DIS) and quasichemical (QUAC) interchange coefficients, $C_{hn,l}^{DIS}$ and $C_{hn,l}^{QUAC}$ ($l$ = 1, Gibbs energy; $l$ = 2, enthalpy; $l$ = 3, heat capacity) for (h,n) contacts[a] in $CH_3(CH_2)_{u-1}OH$ + $CH_3(CH_2)_{v-1}CN$ mixtures.

| (u,v) | $C_{hn,1}^{DIS}$ | $C_{hn,2}^{DIS}$ | $C_{hn,3}^{DIS}$ | $C_{hn,1}^{QUAC}$ | $C_{hn,2}^{QUAC}$ | $C_{hn,3}^{QUAC}$ |
|---|---|---|---|---|---|---|
| (1,1) | 8.6 | 14.1 | –1 | –2 | –3 | 3 |
| (2,1) | 6.35 | 15.8 | –1 | –0.2 | –3 | 3 |
| (3,1) | −2 | 6.6 | –1 | 6 | 1 | 3 |
| (4,1) | −1.5 | 8.2 | –1 | 6 | 1 | 3 |
| (8,1) | 3.3 | 13.4 | –1 | 6 | 1 | 3 |
| (9,1) | 5.1 | 17.3 | –1 | 6 | 1 | 3 |
| (10,1) | 7.8 | 17.3 | –1 | 6 | 1 | 3 |
| (1,3) | 1.9 | −6.7 | –1 | 2 | 7.6 | 3 |
| (2,3) | 2.7 | −5.1 | –1 | 2 | 7.6 | 3 |
| (3,3) | −3 | 8 | –1 | 6 | 1 | 3 |
| (4,3) | −2.5 | 8.4 | –1 | 6 | 1 | 3 |
| (5,3) | −1.6 | 9.6 | –1 | 6 | 1 | 3 |
| (6,3) | −0.4 | 10.4 | –1 | 6 | 1 | 3 |
| (7,3) | 1.15 | 12 | –1 | 6 | 1 | 3 |
| (8,3) | 2.8 | 13.4 | –1 | 6 | 1 | 3 |
| (9,3) | 5.1 | 17.3 | –1 | 6 | 1 | 3 |
| (10,3) | 7.8 | 17.3 | –1 | 6 | 1 | 3 |

[a]type h, OH in 1-alkanols, type n, CN in alkanenitriles

TABLE 2

ERAS parameters for 1-alkanol(1) + alkanenitrile(2) mixtures at 298.15 K

| System | $K_{AB}$ | $\Delta h^*_{AB}$ / kJ·mol$^{-1}$ | $\Delta v^*_{AB}$ / cm$^3$·mol$^{-1}$ | $X_{AB}$ / J·cm$^{-3}$ |
|---|---|---|---|---|
| Methanol + acetonitrile | 66 | −14 | −8 | 14 |
| ethanol + acetonitrile | 45 | −14 | −8.6 | 34 |
| 1-propanol + acetonitrile | 20 | −14 | −12 | 71.5 |
| 1-butanol + acetonitrile | 11 | −14 | −16.3 | 88 |
| 1-decanol + acetonitrile | 4 | −14 | −23 | 130 |
| Methanol + butanenitrile | 50 | −14 | −6.5 | 14 |
| ethanol + butanenitrile | 40 | −14 | −10.3 | 14 |
| 1-propanol + butanenitrile | 5.8 | −14 | −16 | 53 |
| 1-butanol + butanenitrile | 3 | −14 | −23 | 54.4 |
| 1-decanol + butanenitrile | 1.9 | −14 | −23 | 64.3 |

[a] $K_{AB}$, association constant of component A with component B; $\Delta h^*_{AB}$, association enthalpy of component A with component B; $\Delta v^*_{AB}$, association volume of component A with component B; $X_{AB}$, physical parameter

TABLE 3

Molar excess Gibbs energies, $G_m^E$ , at equimolar composition and temperature *T* for $CH_3(CH_2)_{u-1}OH + CH_3(CH_2)_{v-1}CN$ mixtures.

| (u,v) | *T*/K | $N^a$ | $G_m^E$ /J·mol$^{-1}$ | | $\sigma_r(P)^b$ | | | Ref. |
|---|---|---|---|---|---|---|---|---|
| | | | Exp. | DQ | Exp. | DQ | UNIF | |
| (1,1) | 328.15 | 13 | 623 | 618 | 0.003 | 0.006 | 0.003 | 68 |
| (2,1) | 293.15 | 13 | 852 | 850 | 0.001 | 0.021 | 0.016 | 69 |
| | 323.15 | 12 | 795 | 773 | 0.003 | 0.017 | 0.019 | 70 |
| | 343.15 | 13 | 727 | 715 | 0.0007 | 0.014 | 0.011 | 69 |
| | 393.15 | 13 | 533 | 567 | 0.0006 | 0.011 | 0.007 | 69 |
| (3,1) | 328.15 | 10 | 859 | 855 | 0.005 | 0.017 | 0.037 | 71 |
| (4,1) | 333.15 | 8 | 908 | 888 | 0.003 | 0.036 | 0.063 | 10 |
| (1,3) | 278.15 | 11 | 759 | 750 | 0.004 | 0.028 | 0.027 | 72 |
| | 298.15 | 11 | 733 | 740 | 0.003 | 0.025 | 0.012 | 72 |
| | 318.15 | 11 | 710 | 721 | 0.001 | 0.025 | 0.009 | 72 |
| (2,3) | 278.15 | 11 | 809 | 808 | 0.004 | 0.011 | 0.027 | 73 |
| | 298.15 | 11 | 774 | 778 | 0.002 | 0.008 | 0.035 | 73 |
| | 323.15 | 11 | 732 | 727 | 0.006 | 0.009 | 0.041 | 73 |
| (3,3) | 278.15 | 10 | 839 | 804 | 0.009 | 0.034 | 0.050 | 74 |
| | 298.15 | 10 | 754 | 756 | 0.005 | 0.020 | 0.025 | 74 |
| | 323.15 | 10 | 702 | 682 | 0.001 | 0.018 | 0.023 | 74 |
| (4,3) | 278.15 | 10 | 852 | 844 | 0.016 | 0.029 | 0.060 | 73 |
| | 298.15 | 10 | 787 | 794 | 0.002 | 0.021 | 0.031 | 73 |
| | 323.15 | 10 | 710 | 718 | 0.002 | 0.021 | 0.019 | 73 |
| (5,3) | 298.15 | 11 | 834 | 836 | 0.010 | 0.047 | 0.086 | 72 |
| (6,3) | 298.15 | 10 | 854 | 969 | 0.033 | 0.070 | 0.130 | 72 |
| (7,3) | 298.15 | 10 | 908 | 918 | 0.014 | 0.050 | 0.120 | 72 |
| (8,3) | 298.15 | 13 | 954 | 943 | 0.019 | 0.044 | 0.130 | 72 |
| (10,3) | 298.15 | 9 | 1049 | 1045 | 0.007 | 0.077 | 0.180 | 72 |

[a]number of data points; [b]eq. (12)

TABLE 4

Molar excess enthalpies, $H_m^E$, at equimolar composition and temperature *T* for $CH_3(CH_2)_{u-1}OH$ + $CH_3(CH_2)_{v-1}CN$ mixtures.

| (u,v) | *T*/K | *N*[a] | $H_m^E$/J·mol$^{-1}$ | | $dev(H_m^E)$ [b] | | | Ref. |
|---|---|---|---|---|---|---|---|---|
| | | | Exp. | DQ | Exp. | DQ | UNIF/ERAS | |
| (1,1) | 298.15 | 14 | 1086 | 1104 | 0.001 | 0.035 | 0.013/0.065 | 5 |
| | | 9 | 1113 | | | 0.029 | 0.022 | 75 |
| | | 23 | 1102 | | 0.005 | 0.030 | 0.023 | 76 |
| | 308.15 | 14 | 1140 | 1152 | 0.001 | 0.038 | 0.018 | 5 |
| (2,1) | 298.15 | 13 | 1500 | 1542 | 0.002 | 0.025 | 0.028/0.087 | 5 |
| | | 20 | 1496 | | 0.003 | 0.025 | 0.031 | 77 |
| | 308.15 | 14 | 1574 | 1602 | 0.001 | 0.025 | 0.034 | 5 |
| | | 20 | 1571 | | 0.006 | 0.025 | 0.040 | 77 |
| | 318.15 | 20 | 1645 | 1654 | 0.005 | 0.022 | 0.073 | 77 |
| (3,1) | 298.15 | 20 | 1827 | 1841 | 0.002 | 0.022 | 0.022/0.105 | 6 |
| | | 24 | 1844 | | 0.007 | 0.026 | 0.019 | 78 |
| | 308.15 | 22 | 1921 | 1919 | 0.003 | 0.020 | 0.017 | 78 |
| | 318.15 | 22 | 1983 | 1995 | 0.002 | 0.013 | 0.043 | 78 |
| (4,1) | 298.15 | 17 | 2037 | 2075 | 0.004 | 0.012 | 0.054/0.122 | 7 |
| | | 9 | 2067 | | 0.0006 | 0.017 | 0.057 | 75 |
| | | 25 | 2090 | | 0.007 | 0.009 | 0.045 | 76 |
| (9,1) | 298.15 | 20 | 2717 | 2678 | 0.0003 | 0.055 | 0.171 | 58 |
| | 308.15 | 20 | 3050 | 2789 | 0.007 | 0.066 | 0.057 | 58 |
| (10,1) | 298.15 | 20 | 3032 | 2773 | 0.0005 | 0.049 | 0.106/0.202 | 58 |
| | 308.15 | 20 | 3297 | 2883 | 0.009 | 0.094 | 0.055 | 58 |
| (1,3) | 298.15 | 17 | 979 | 953 | 0.007 | 0.026 | 0.169/0.046 | 72 |
| | | 11 | 936 | | | 0.072 | 0.143 | 14 |
| (2,3) | 298.15 | 11 | 1319 | 1305 | 0.006 | 0.059 | 0.186/0.051 | 73 |
| | | 12 | 1185 | | 0.003 | 0.118 | 0.143 | 14 |
| (3,3) | 298.15 | 12 | 1513 | 1541 | 0.011 | 0.039 | 0.229/0.090 | 74 |
| | | 12 | 1389 | | 0.006 | 0.129 | 0.180 | 14 |
| (4,3) | 298.15 | 12 | 1589 | 1608 | 0.005 | 0.027 | 0.201/0.112 | 73 |
| (5,3) | 298.15 | 14 | 1707 | 1732 | 0.010 | 0.016 | 0.237 | 72 |
| (6,3) | 298.15 | 19 | 1767 | 1805 | 0.003 | 0.022 | 0.208 | 79 |
| (7,3) | 298.15 | 13 | 1867 | 1906 | 0.009 | 0.018 | 0.208 | 72 |
| (8,3) | 298.15 | 11 | 1970 | 1985 | 0.006 | 0.025 | 0.192 | 79 |

TABLE 4 (continued)

| | | | | | | | | |
|---|---|---|---|---|---|---|---|---|
| (9,3) | 298.15 | 13 | 2018 | 2109 | 0.005 | 0.030 | 0.201 | 72 |
| (10,3) | 298.15 | 18 | 2171 | 2170 | 0.004 | 0.011 | 0.118/0.111 | 72 |

[a]number of data points; [b]eq. (13)

TABLE 5

Solid-liquid equilibria for $CH_3(CH_2)_{u-1}OH + CH_3(CH_2)_{v-1}CN$ mixtures [15].

| (u,v) | $N^a$ | $\Delta(T)^b$/K | $\sigma_r(T)^c$ |
|---|---|---|---|
| (8,1) | 34 | 1.1 | 0.006 |
| (9,1) | 32 | 1.1 | 0.006 |
| (10,1) | 26 | 1.3 | 0.005 |
| (8,2) | 22 | 1.1 | 0.007 |
| (9,2) | 23 | 0.98 | 0.005 |
| (10,2) | 18 | 0.96 | 0.004 |
| (8,3) | 28 | 0.9 | 0.006 |
| (9,3) | 30 | 1.3 | 0.010 |
| (10,3) | 27 | 1.5 | 0.007 |

[a]number of data points; [b]eq. (15); [c]eq. (14)

TABLE 6

Coordinates of the critical points, composition ( $x_{1c}$ ) and temperature ( $T_c$ ) for 1-alkanol(1) + acetonitrile(2) mixtures [15]

| 1-alkanol | $x_{1c}$ | | $T_c$/K | |
|---|---|---|---|---|
| | Exp. | DQ. | Exp. | DQ |
| 1-octanol | 0.271 | 0.250 | 282.07 | 284.8 |
| 1-decanol | 0.244 | 0.193 | 297.05 | 301.7 |

TABLE 7

Partial excess molar enthalpies at infinite dilution, $H_{mi}^{E,\infty}$, of 1-alkanol(1) + alkanenitrile(2) mixtures at 298.15 K.

| System | $H_{m1}^{E,\infty}$ /kJ·mol$^{-1}$ | | $H_{m2}^{E,\infty}$ /kJ·mol$^{-1}$ | | Ref. |
|---|---|---|---|---|---|
| | Exp. | DISQUAC | Exp. | DISQUAC | |
| Methanol(1) + acetonitrile(2) | 5.66 | 5.37 | 4.34 | 5.19 | 80 |
| | 6.1[a] | | 4.1[a] | | 5 |
| Methanol(1) + butanenitrile(2) | | | 3.81 | 4.68 | 80 |
| | 4.9[a] | 5.93 | 3.6[a] | | 72 |
| Ethanol(1) + acetonitrile(2) | 7.88 | 7.37 | | 5.26 | 80 |
| | 8.0[a] | | 6.3[a] | | 5 |
| 1-propanol(1) + acetonitrile(2) | 8.84 | 9.47 | | 10.16 | 80 |
| | 9.2[a] | | 8.1[a] | | 6 |
| 1-butanol(1) + acetonitrile(2) | 10.29 | 11.21 | | 10.55 | 80 |
| | 10.4[a] | | 9.2[a] | | 7 |

[a]value obtained from $H_{m}^{E}$ data over the whole concentration range

TABLE 8

Molar excess volumes, $V_m^E$, at equimolar composition and 298.15 K for 1-alkanol + alkanenitrile mixtures.

| System | $V_m^E$ /cm$^3$·mol$^{-1}$ | | Ref. |
|---|---|---|---|
| | Exp. | ERAS | |
| methanol + acetonitrile | −0.1424 | −0.148 | 57 |
| ethanol + acetonitrile | −0.0258 | −0.027 | 57 |
| 1-propanol + acetonitrile | 0.0544 | 0.057 | 57 |
| 1-butanol + acetonitrile | 0.1045 | 0.110 | 57 |
| 1-decanol + acetonitrile | 0.340 | 0.331 | 59 |
| methanol + butanenitrile | −0.0124 | −0.014 | 72 |
| ethanol + butanenitrile | 0.0573 | 0.066 | 73 |
| 1-propanol + butanenitrile | 0.0796 | 0.065 | 74 |
| 1-butanol + butanenitrile | 0.1044 | 0.110 | 73 |
| 1-decanol + butanenitrile | 0.3244 | 0.328 | 72 |

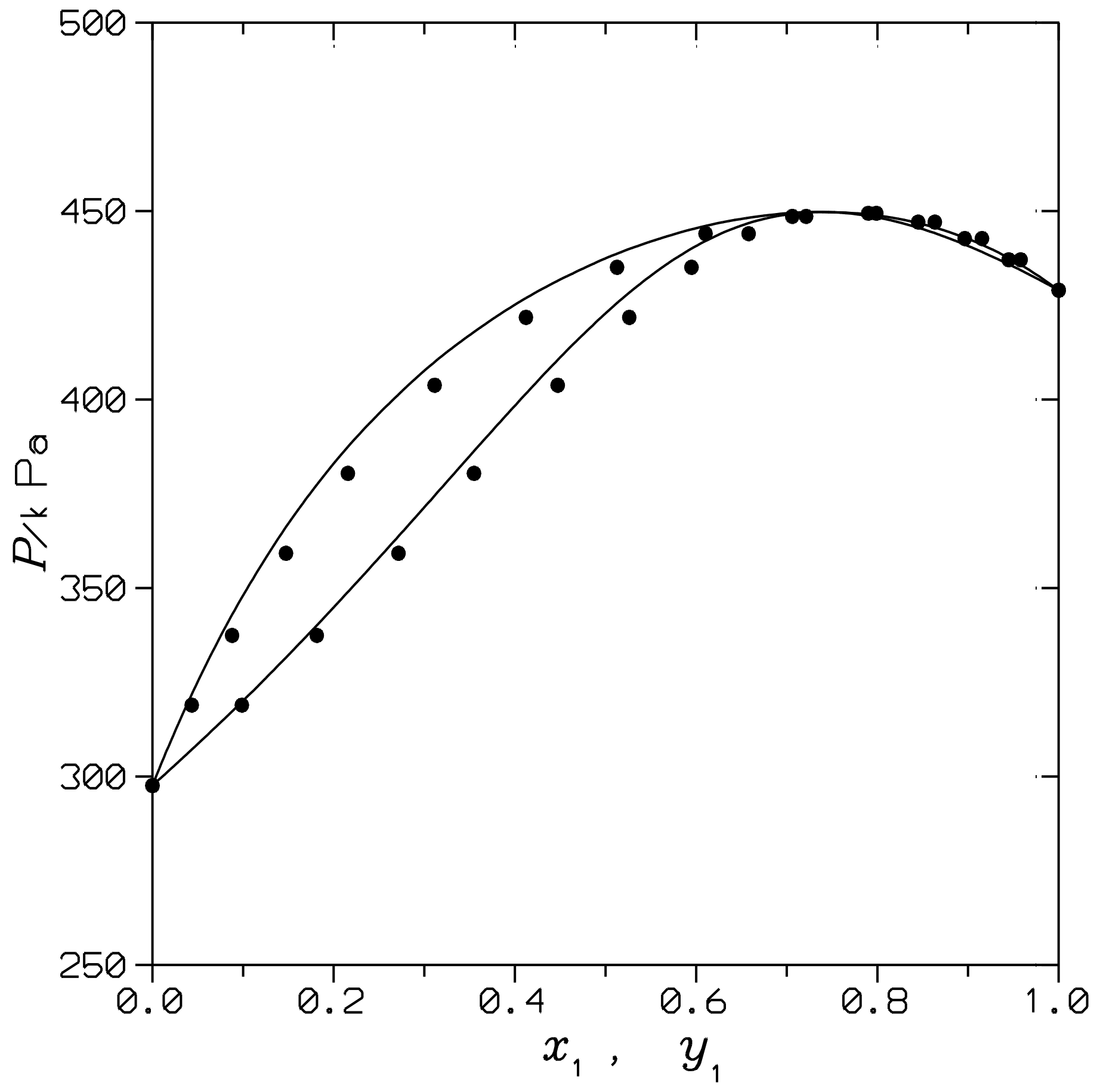

**Figure 1**. VLE phase diagram for the ethanol(1) + acetonitrile(2) system at 393.15 K. Points, experimental results [69]. Solid lines DISQUAC calculations with interaction parameters listed in Table 1.

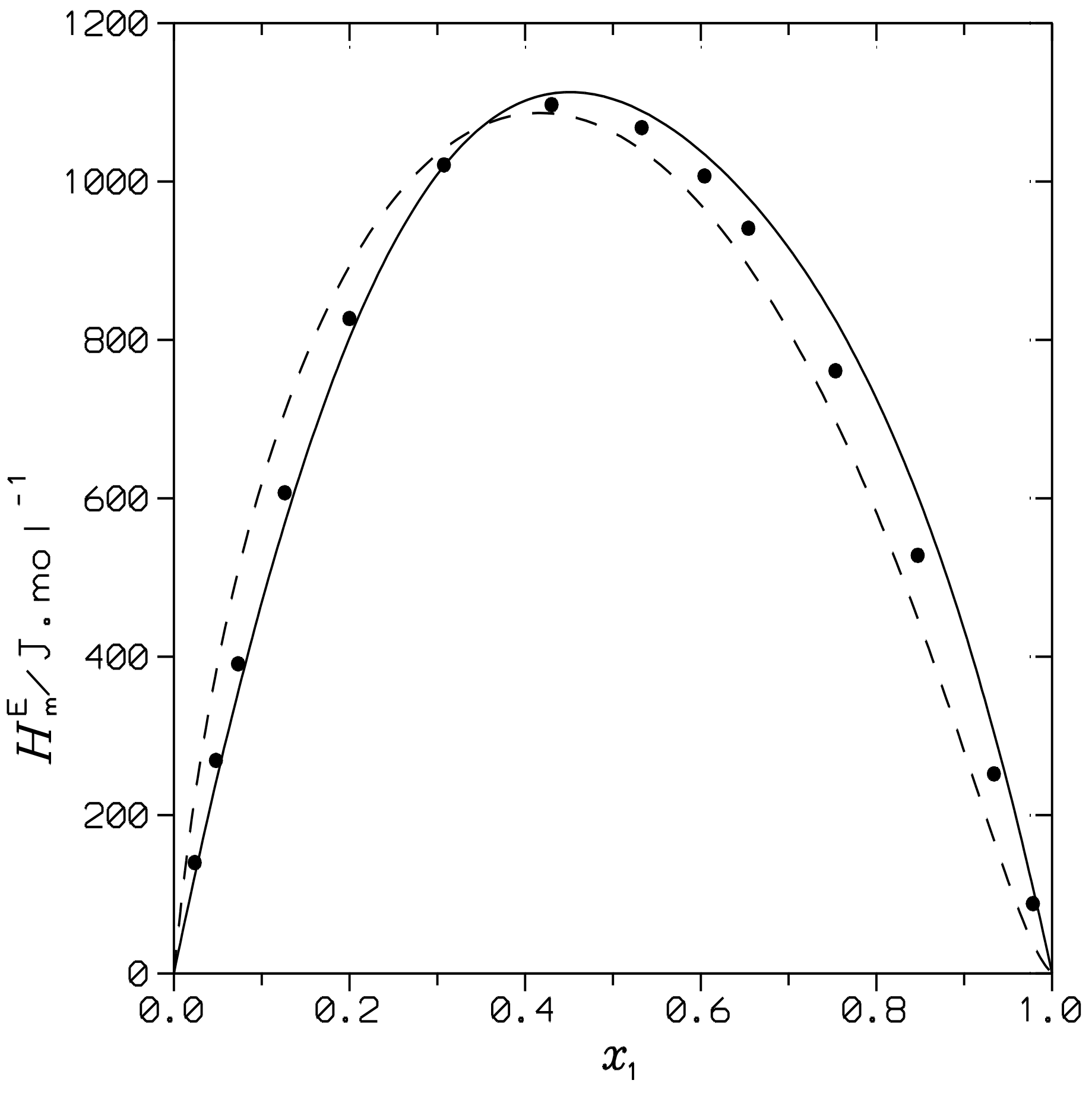

**Figure 2**. $H_m^E$ for the methanol(1) + acetonitrile(2) system at 298.15 K. Points, experimental results [5]. Solid line, DISQUAC calculations with interaction parameters listed in Table 1. Dashed line, ERAS results with parameters collected in Table 2

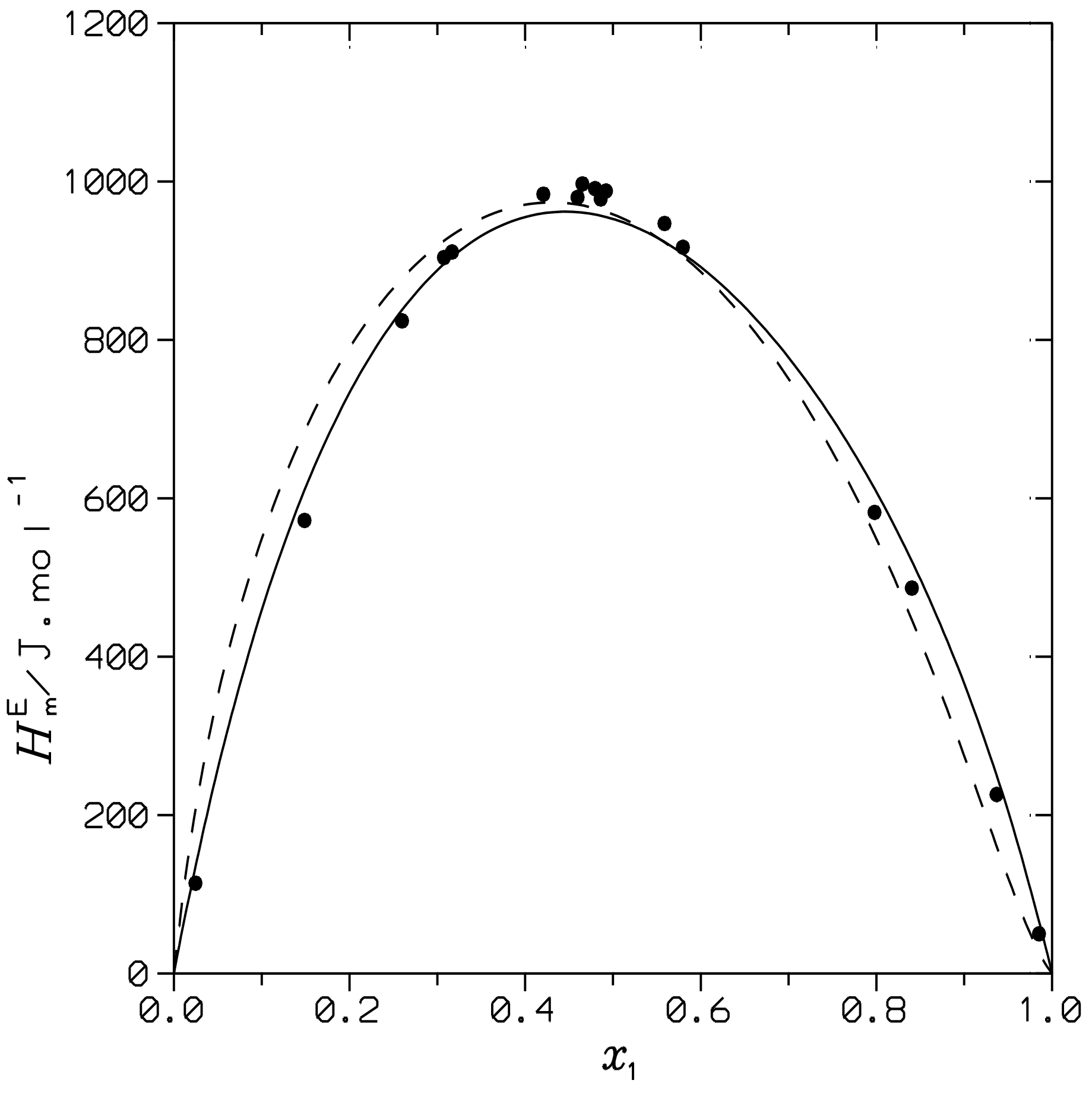

**Figure 3**. $H_m^E$ for the methanol(1) + butanenitrile(2) system at 298.15 K. Points, experimental results [72]. Solid line, DISQUAC calculations with interaction parameters listed in Table 1. Dashed line, ERAS results with parameters collected in Table 2.

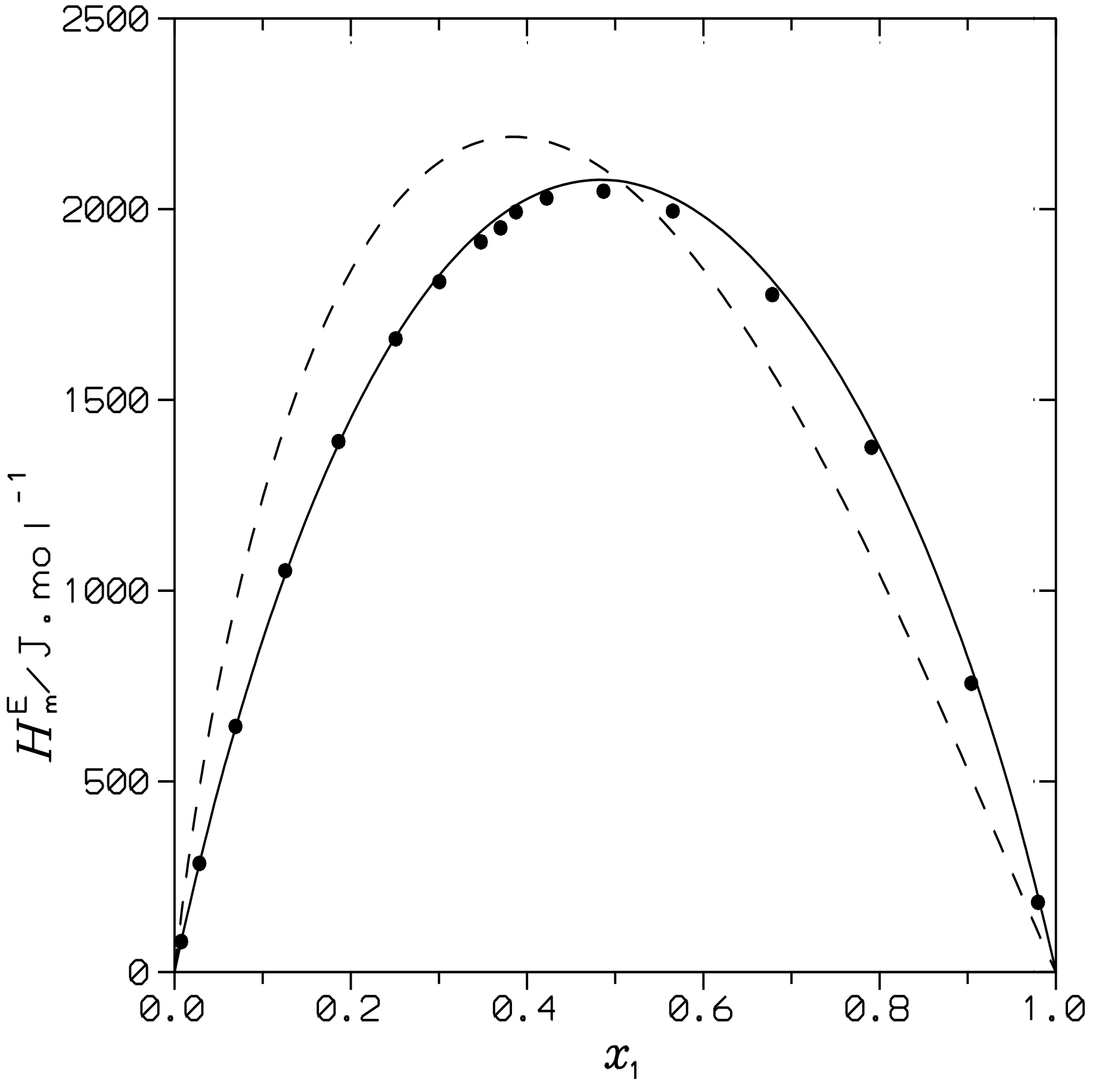

**Figure 4**. $H_m^E$ for the 1-butanol(1) + acetonitrile(2) system at 298.15 K. Points, experimental results [7]. Solid line, DISQUAC calculations with interaction parameters listed in Table 1. Dashed line, ERAS results with parameters collected in Table 2

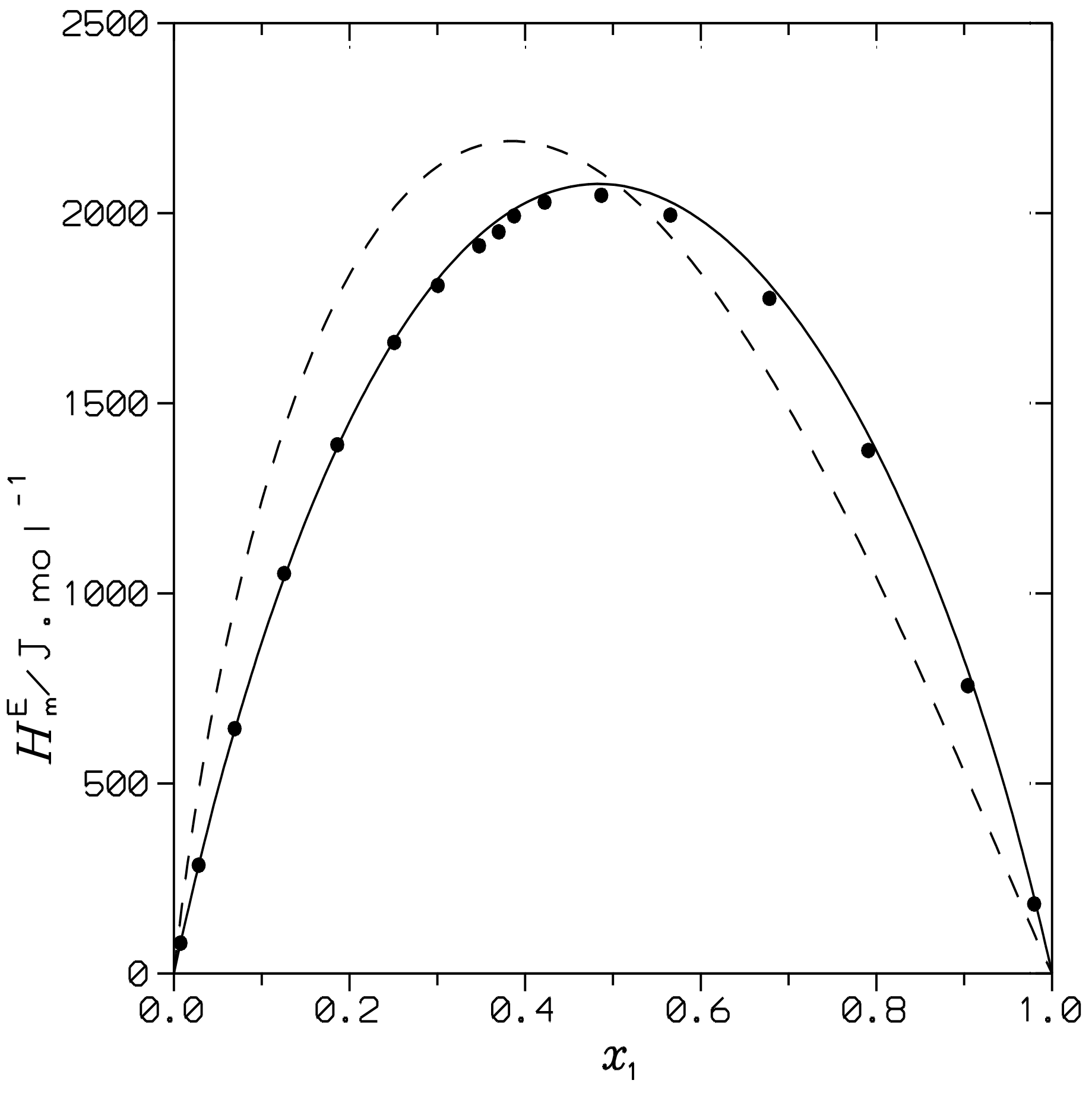

**Figure 5** $H_m^E$ for the 1-decanol(1) + butanenitrile(2) system at 298.15 K. Points, experimental results [72]. Solid line, DISQUAC calculations with interaction parameters listed in Table 1. Dashed line, ERAS results with parameters collected in Table 2

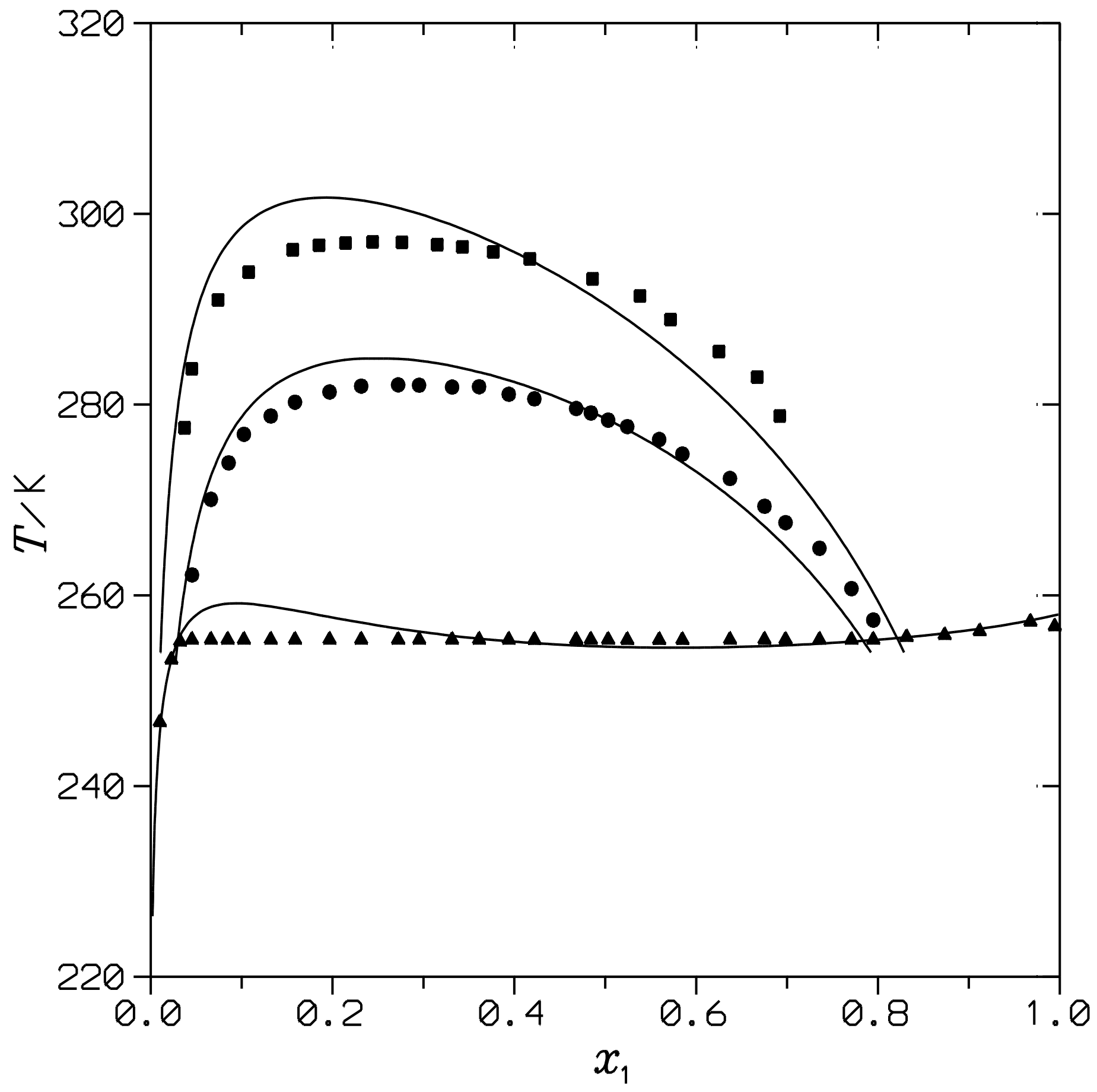

**Figure 6**. LLE and SLE phase diagrams for 1-alkanol(1) + acetonitrile(2) system. Points, experimental results [15]: (●), LLE for the 1-octanol mixture; (■), LLE for the 1-decanol solution; (▲), SLE for the 1-octanol system. Solid lines DISQUAC calculations with interaction parameters listed in Table 1.